\documentclass[prl,twocolumn,nofootinbib,superscriptaddress,showpacs,showkeys]{revtex4-1}
\usepackage{amssymb}
\usepackage{amsmath}
\usepackage{amstext}
\usepackage{amsfonts}
\usepackage{bbold}
\usepackage{bm}
\usepackage{graphics}
\usepackage{mathtools}
\usepackage{lmodern}
\usepackage{graphicx}
\usepackage[capitalise]{cleveref}
\usepackage{xspace}
\usepackage{url}

\newcommand{\XYZ}{$XY\!Z$\xspace}

\begin{document}

\title{A $bb\bar b\bar b$ di-bottomonium at the LHC?} 

\author{Angelo~Esposito}
\affiliation{Department of Physics, Center for Theoretical Physics, Columbia University, 538W 120th Street, New York, NY, 10027, USA}
\affiliation{INFN, Sezione di Roma, Piazzale A. Moro 2, I-00185 Rome, Italy}
\affiliation{Theoretical Particle Physics Laboratory, Institute of Physics, EPFL, Lausanne, Switzerland}
\author{Antonio~D.~Polosa}
\affiliation{Dipartimento di Fisica and INFN, Sapienza Universit\`a di Roma, P.le Aldo Moro 2, I-00185 Roma, Italy}

\begin{abstract}
We study the case of a di-bottomonium $bb\bar b\bar b$ particle at the LHC. 
Its mass and decay width are obtained within a new model of diquark-antidiquark interactions in tetraquarks.
The same model solves several open problems on the phenomenology of the experimentally better  known $X,Z$ states.  We show that the $bb\bar b\bar b$ tetraquark is expected approximately 100 MeV below threshold, and compare to a recent analysis by LHCb seeking it in the $\Upsilon\mu\mu$ final state.
\end{abstract}


\maketitle


\section{Introduction}
Very recently the LHCb collaboration presented an intriguing analysis seeking the  $bb\bar b\bar b$ exotic meson in the $\Upsilon\mu\mu$ final state~\cite{Aaij:2018zrb}. No observation was made. This seems to contradict some preliminary results (not yet approved) from the CMS experiment, claiming an indication for the existence of such particle~\cite{slides}, with a global significance of $3.6\sigma$ and a mass around $18.4$~GeV. This would set it at 500 MeV \emph{below} the $\Upsilon\Upsilon$ threshold. 

The possible existence of a $bb\bar b\bar b$ state has already been considered in a fair number of theoretical papers~\cite{heller,berezhnoy,wu,chen,Karliner:2016zzc,Bai:2016int,Anwar:2017toa,Eichten:2017ual,Hughes:2017xie,Wang:2017jtz}. Some of these works pointed out that  the attraction in the $bb$ diquark may be so strong as to make the double $b$ tetraquark  stable under strong interaction decays. 
Double charm and beauty tetraquarks have been considered in~\cite{Buccella:2006fn,Esposito:2013fma,Guerrieri:2014nxa} and, more recently, in~\cite{Karliner:2017qjm} and~\cite{Eichten:2017ffp,Eichten:2017ual,Richard:2018yrm,Wang:2017dtg}. 
The presence of tetraquarks in the double charm or beauty channel is also indicated in a number of non-perturbative approaches such as the Heavy Quark-Diquark Symmetry~\cite{Manohar:2000dt,Savage:1990di,Mehen:2017nrh}, lattice QCD --- see e.g.~\cite{Bicudo:2016ooe,Hughes:2017xie} and references therein --- and the Born-Oppenheimer approximation, see~\cite{Bicudo:2017szl} and references therein.

The observation/absence of this exotic resonance is very informative on the nature of the so-called \XYZ states,  made of four quarks, as observed at lepton and hadron colliders in the past  decade~\cite{Esposito:2016noz,Ali:2017jda,Lebed:2016hpi,Guo:2017jvc}. 

The debate about the internal structure of these particles has been quite lively and cannot  be considered as settled yet. We believe that, rather than being kinematical effects~\cite{Bugg:2004rk} or meson molecules~\cite{Guo:2017jvc}, most of these objects are compact tetraquark states~\cite{maiani}.

In this paper we estimate the mass and the total width of an hypothetical $bb\bar b \bar b$ state in a new compact tetraquark model, which we use to understand the properties of the best known $X$ and $Z$ resonances.  

We conclude that the $bb\bar b\bar b$ state that we predict could hardly be detected by LHCb, whereas  it it is unlikely for it to be observed in a $\Upsilon\mu\mu$ final state with the current CMS sensitivity (tens of pb~\cite{Khachatryan:2016ydm}). We make an assumption on its prompt production cross section and comment on how its detectability depends on the main parameter of  the model presented here.


\section{Background facts on $X,Z$ resonances}
Soon after the first attempts to understand  the \XYZ states in the framework of the constituent quark model and SU(3), it was realized that a description in terms of diquark-antidiquark states could be rather efficient at reproducing their spectra, provided that spin-spin interactions among quarks are confined within the diquarks~\cite{maiani}. 

Moreover, the $X$ and $Z$ resonances, made of two heavy and two light quarks, decay more promptly into open charm/beauty meson, rather than into quarkonia. 

Both these aspects cannot be immediately understood  from the viewpoint of the constituent quark model.

It was also shown that the correct mass splitting between radial excitations of tetraquark states is reproduced well assuming that the diquarkanium potential is purely linear with distance~\cite{Chen:2015dig}, with little role of the Coulomb term
at short distances.

\begin{figure}
 \centering
 \includegraphics[width=0.45\textwidth]{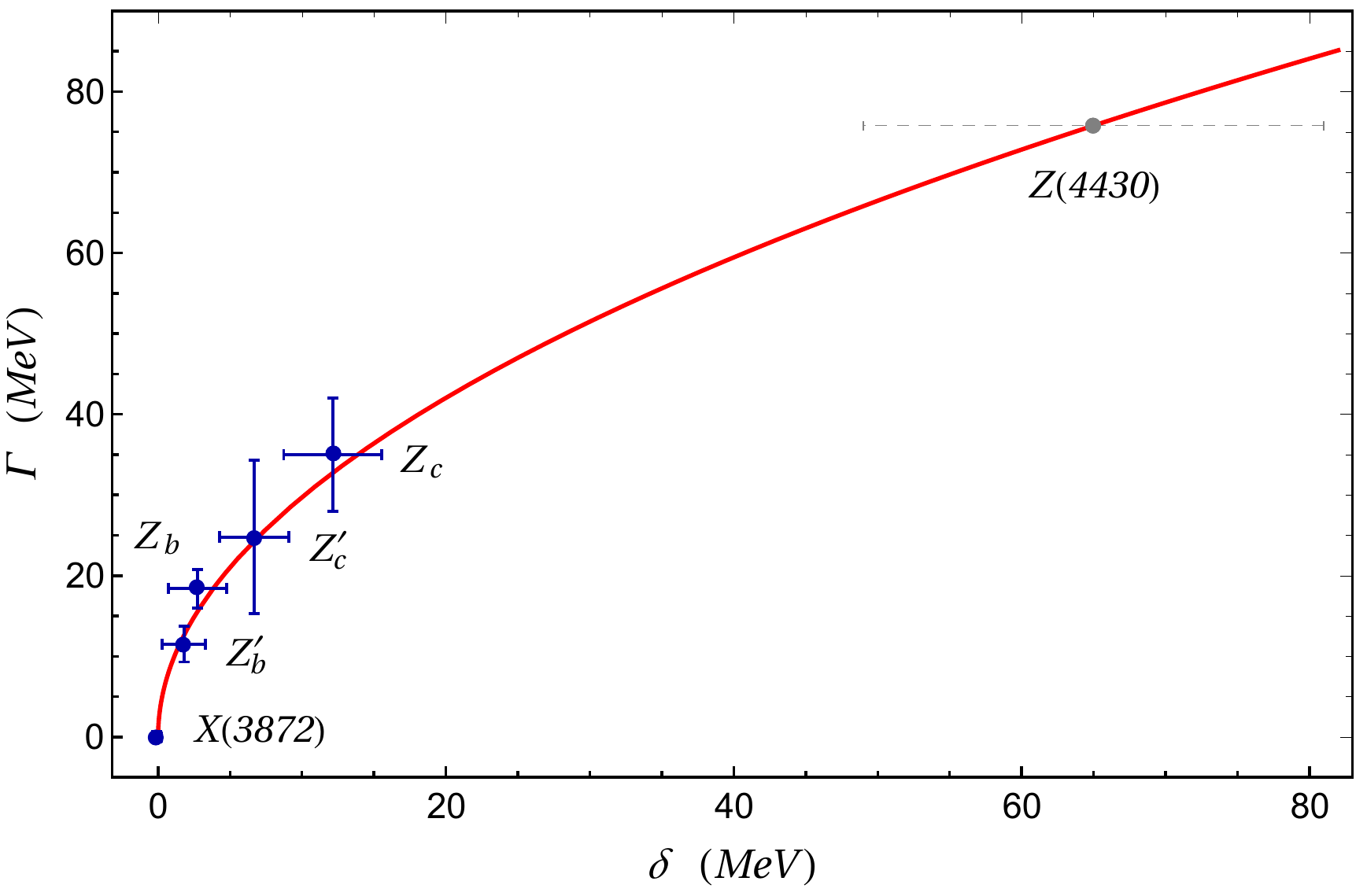}
 \caption{Comparison between the experimental values of $\delta$ and $\Gamma$ for the $XZ$ resonances and the $\Gamma=A\sqrt{\delta}$ law. All the ground state tetraquarks fit the curve remarkably well, with a value of the fitted parameter $A=9.4\pm1.5$~MeV$^{1/2}$~\cite{Esposito:2016itg} common to \emph{both} the charm and beauty resonances. We also report the expected with of the $Z(4430)$, which is underestimated with respect to the observed one. Indeed, being the $Z(4430)$ an excited state, the decay towards a ground state tetraquark (rather than a meson pair) is also available.} \label{fig:fit}
\end{figure}

Another striking fact was observed in~\cite{Esposito:2016itg}. The total widths of the $X$ and $Z$ resonances seem to follow rather well a $\Gamma=A\sqrt{\delta}$ behavior,
where $\delta>0$ is the mass difference between the tetraquark and the closer, from below, meson-meson threshold having the same quantum numbers. The $X(3872)$ and the four resonances $Z^{(\prime)}_c, Z_b^{(\prime)}$ are all found, experimentally {\it above} their related meson-meson thresholds ($D^0\bar D^{*0}$ for the $X$, etc.). 
The interesting point, not fully appreciated in~\cite{Esposito:2016itg}, is that the constant  $A\simeq10$~MeV$^{1/2}$ fitted from data, is the same for charmed and beauty resonances (see Fig.~\ref{fig:fit}) --- something that cannot be reconciled with simple phase space arguments.  

In this paper we solve all the the previous issues with an assumption on the inter-diquark potential: that the diquark-antidiquark pair is separated by a repulsive barrier. 
To the best of our knowledge this possibility was first sketched  in~\cite{Selem:2006nd} and further considered in~\cite{Maiani:2017kyi}.

In our approach the suppression of the contact spin-spin interactions outside the diquarks is simply due to their spatial separation. The preferred decay into open flavor mesons rather than quarkonia is also  understood: the barrier tunneling of heavy quarks is suppressed with respect to that of light quarks. Moreover, the marginal role of the Coulomb term in the diquarkonium potential, discussed in~\cite{Chen:2015dig}, is explained by the fact that, at short distances, the potential is dominated by the repulsive  barrier. 

Finally, this picture allows to compute $A$ (in  $\Gamma=A\sqrt{\delta}$) for very reasonable parameters of the model and to explain under which conditions it turns out to have the same value for both the charm and beauty resonances. This is obtained  if the \emph{size} of the tetraquark in the beauty sector, is slightly smaller  than that in the charm one, which  agrees with the standard quarkonium picture.

This idea has also important consequences on the decay properties of the charged partners of the $X$~\cite{Maiani:2017kyi}, making them particularly elusive.

\section{Quasi-compact diquark-antidiquark states}
When three quarks are produced in a small enough phase space volume, there is only one way to form a color singlet, i.e. a baryon, $\epsilon_{ijk}q^iq^jq^k$.  Similarly,  a quark-antiquark pair  can neutralize the color in only one way, as in a meson, $q^i\bar q_i$.   More alternatives arise for a number of quarks/antiquarks larger than three.  In the case of two quarks and two antiquarks, one can form two mesons or a color neutral diquark-antidiquark pair, $[q_1q_2]_i[\bar q_3\bar q_4]^i$.

In general, the quantum state describing two quarks and two antiquarks is a statistical admixture of these two alternatives, with  probabilities which cannot be computed from first principles. 

A  special case, particularly relevant in phenomenology, is that of $Q\bar Q q\bar q^{\,\prime}$ where $Q=c,b$ and $q^{(\prime)}=u,d,s$. 

The energy stored in (the center of mass) of this four quark system can be distributed to form a 
diquark-antidiquark bound state at rest, with mass
\begin{align} \label{eq:mass}
E=\widetilde m_1 +\widetilde m_2-B+H_I\,,
\end{align}
where $B>0$ is (minus) the binding energy between the diquarks of masses $\widetilde m_{1,2}$, and $H_I$ is the spin-spin Hamiltonian with interactions confined within the diquarks~\cite{maiani}:
\begin{align}
H_I=2\kappa_{Qq}\left(\bm{S}_Q\cdot\bm{S}_q+\bm{S}_{\bar Q}\cdot\bm{S}_{\bar q^\prime}\right)\,.
\end{align}

We assume however that, unless prevented by some mechanism, this state would immediately fall apart into a meson-meson pair  with same energy
\begin{align} \label{delta}
E=m_1+m_2+\delta\,,
\end{align}
where $m_{1,2}$ are the final meson masses. If instead two mesons are directly formed out of  the four quark system,  the same two free mesons will eventually be observed, without forming any molecule.
This is even more true if the two mesons recoil with sufficiently high reletive momenta, as discussed in a number of papers --- see~\cite{Bignamini:2009sk,Artoisenet:2009wk,Bignamini:2009fn,Esposito:2013ada} and~\cite{Esposito:2015fsa} for a comparison to data.

For example, a compact spin one and positive parity tetraquark like the $X(3872)$,  should simply fall apart into a free $D\bar D^*$ or  $J/\psi\,\rho$ pair by spontaneously rearranging its internal color configuration.

From what just said,  there is no space for narrow resonances with the quantum numbers of four quarks.

\begin{figure}[t]
\centering
   \includegraphics[width=0.43\textwidth]{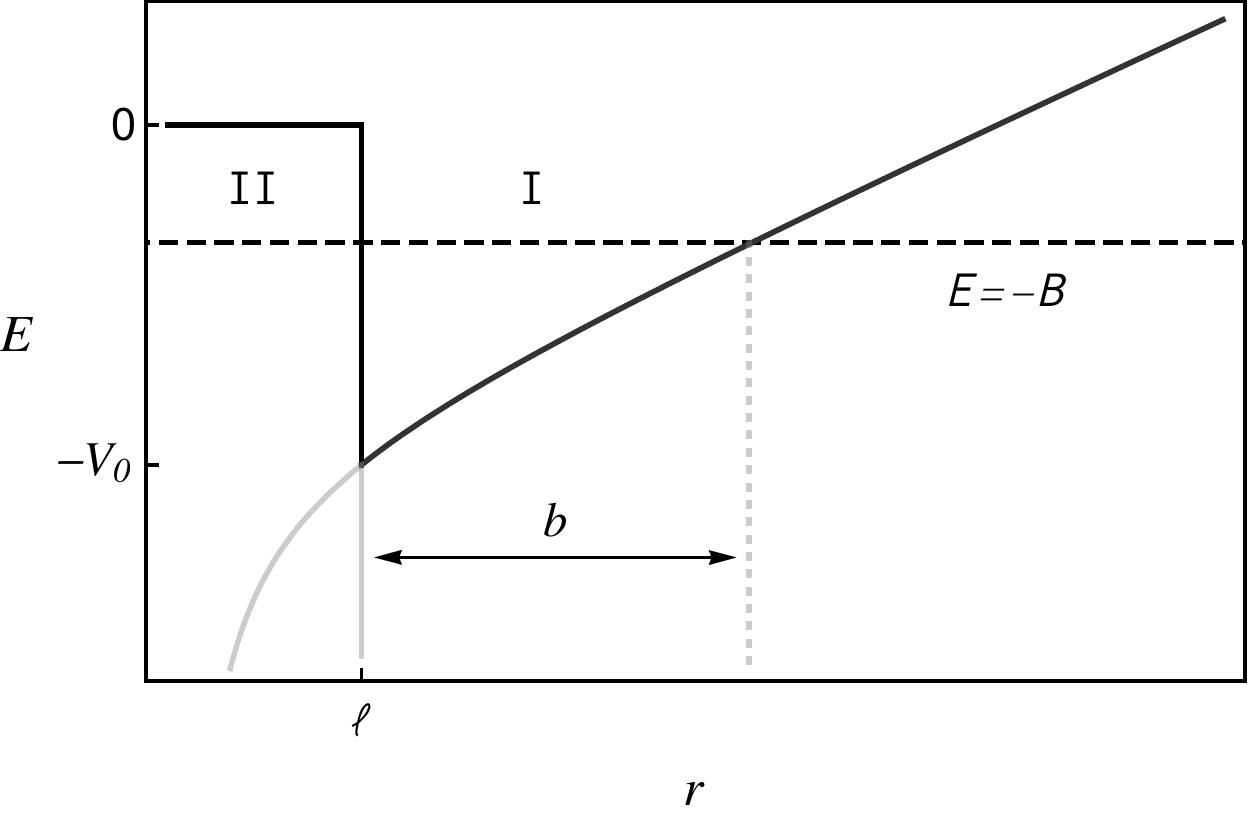}
\caption{\small Potential between diquarks as a function of their distance. Here $b=(V_0-B)/\nu$ is the classical turning point. The diquark-antidiquark pair is produced in region I. If a quark tunnels through the barrier or if  $E>0$, color can be rearranged and the state decays into a meson pair. \label{frame0}}
\end{figure}

There is another possibility though.
As commented in the Introduction, the diquarks could  be produced at some large enough relative distance from each other --- region I in Fig.~\ref{frame0}. A potential barrier between diquarks could make the tetraquark metastable against collapse and fall apart decay, which happens if one of the quarks tunnels towards the other side. (Provided that the diquark-antidiquark system has a binding energy below the height of the barrier.) This suppression of the decay into pairs of mesons is also envisaged in the $1/N$ QCD expansion~\cite{Weinberg:2013cfa,Knecht:2013yqa,Cohen:2014via,Maiani:2016hxw,Maiani:2018pef}.

When close in space, the two diquarks see each other as color dipoles rather than point-like sources.
On the one hand, the attraction between quarks and antiquarks tends to disintegrate the diquarks. On the other hand, in order to do that, the diquark binding energy must be overcome. These effects increase when the separation decreases and produce a repulsion among diquark and antidiquark, i.e. a component in the potential increasing at decreasing distance. If this effect wins against the decrease due to the color attraction, it will produce the barrier in region II of Fig.~\ref{frame0}.
When the diquark-antidiquark pair is instead separated by a larger distance, a strong attraction is felt between the point-like $\bar{\bm 3}$ and $\bm 3$ color sources~\footnote{
A tetraquark could also be thought as a bound state of a $\bm 6$ diquark  and a $\bar{\bm 6}$ antidiquark. However if the $\bm 6,\bar{\bm 6}$ color channels are repulsive, as in the one-gluon-exchange,  the argument presented above is not applicable. There is no binding energy to pay to disintegrate the diquark. We might then simply conclude that $\bm 6$-$\bar{\bm 6}$ states also convert freely into pairs of mesons.}.

Note that the potential barrier also suppresses the mixing of the tetraquark with a molecular state, which would require the quark-antiquark pair to be close in space.

\section{A simple model for the diquark-antidiquark repulsion}
In what follows we propose a simple model to work with the  conjecture of  diquarks segregated in space. As already anticipated, we assume that the inter-diquark potential is given by a repulsive barrier of width $\ell$, beyond which it takes the form
\begin{align}
V(r)=\nu (r-\ell)-V_0 \qquad \text{ for }\, r>\ell\,,
\label{potenziale}
\end{align}
with no Coulomb term, and $\nu=0.25$~GeV$^2$, as obtained from $2S-1S$ and $2P-1P$ mass splittings in the charm sector~\cite{Chen:2015dig}. The absence of the Coulomb term is interpreted to be due to the dominance of the barrier at short distances, see Fig.~\ref{frame0}.

The potential is defined up to a constant $V_0$, which we take in a way that allows to express the  mass of the states in the form of Eq.~\eqref{eq:mass}, with $B$ taken (negatively) from $E=0$~\footnote{If $V_0$ were set to zero, then Eq.~\eqref{tt} would contain the difference between the binding energy and the top of the barrier.}.
Given that the spin-spin correction for the $[cq]$ diquark is $\kappa_{cq}=67~\mathrm{MeV}$~\cite{Esposito:2016noz},
to reproduce both the  masses~\footnote{$M(X(3872))=M(1S)-\kappa_{cq}$ and $M(Z(4430))=M(2S)-\kappa_{cq}$ where $M(1S),M(2S),M(2P),\dots $ are related to the eigenvalues  of the Schr\"odinger equation with the potential~\eqref{potenziale}, e.g.  $M(1S)=2\widetilde m_c-B_c(1S)$ --- see~\eqref{eq:mass}. We refer to $B_{c,b}$ as to the binding energies in the $c,b$ systems.} of the $X(3872)$ and the $Z(4430)$, we need $V_0^{(c)}=0.76$~GeV, giving a binding energy $B_c=0.021$~GeV. This represents the negative gap with respect to the height of the barrier. Above the barrier the diquark-antidiquark system falls apart into an open charm/beauty  pair. 

In the beauty sector the observed mass splitting between the first radial excitation of the bottomonium and its ground state is almost the same as in the charm sector. We expect this to occur in diquarkonia as well. This leads to retain the same value for $\nu$ as given above. In order to get the right mass for the $Z_b(10610)$, we need
\begin{align} \label{eq:binding}
B_b\simeq B_c\,,
\end{align}
which fixes $V_0^{(b)}=0.55$~GeV. There are no data on radial excitations, so this is the best we can do with parameters and data at hand. 

Given the potential~\eqref{potenziale}, a bound state at $E=-B$ is allowed if
\begin{align}
\int_0^bdr\,p(r)=\left(n+\frac{3}{4}\right)\pi\,,\qquad b=(V_0-B)/\nu
\end{align}
with $n=0,1,2,\dots$ and  $p(r)=\sqrt{2m(E-V(r))}$. 
In the charm case for $n=0$ we find $B_c=0.026$~GeV, which  is indeed in very good agreement to what needed to reproduce the mass of the $X(3872)$ --- see above.

The average distance of the diquark from the barrier will then be $\bar R\simeq b/2$, and the total radius of the system
\begin{align} \label{stima}
R\simeq\bar R+\ell=\frac{V_0-B}{2\nu}+\ell.
\end{align}


\section{Tetraquark life-time: the $\delta>0$ case}
Consider the one-to-two decay $\alpha \to \beta$ in which the initial state $\alpha$ is the diquark-antidiquark pair with a spatial separation as in region  I, whereas $\beta$ is the final free meson pair.  The barrier in II is what makes the diquark-antidiquark system metastable, as observed.  

The matrix element of the $\alpha\to \beta$ decay is given by the barrier transparency $\mathcal T$ at II.
In this picture, the rate of the  $\alpha\to \beta$ process corresponds essentially  to  the observed total width of the $X, Z$ resonances~\cite{Esposito:2016noz}.

The differential rate is~\cite{fermi}
\begin{align} \label{1to2}
 \frac{d\Gamma}{d\Omega}(\alpha\to \beta)=\frac{2\pi}{R}|M_{\beta\alpha}|^2\,\frac{k^\prime E_1^\prime E_2^\prime}{E}\,,
 \end{align}
where  $R$ has been estimated in the previous Section. Here 
\begin{align*}
k^\prime=\frac{\sqrt{\lambda(E^2,m_1^2,m_2^2)}}{2E}\,,\qquad E_{1,2}^\prime=\frac{E^2-m_{2,1}^2+m_{1,2}^2}{2E}\,,
\end{align*}
and $\lambda(x,y,z)$ is the K\"all\'en triangular function. 

In our model the matrix element is given by
\begin{align}
|M_{\beta\alpha}|^2=\frac{\mathcal{T}}{2\widetilde m (m_1+m_2)}\,.
\end{align}
The denominator comes from the normalization of the incoming and outgoing states. The barrier transparency factor is (the current density beyond the barrier)
\begin{align}
 \mathcal T= \exp\left(-2\ell\sqrt{2M B}\right)\label{tt}\,,
\end{align}
where $M$ is the mass of the tunneling particle and $\ell$ the barrier width in Fig.~\ref{frame0}, region II.

The light quarks in the diquarks can tunnel through the barrier and hence produce the decay of the tetraquark. When that happens, nothing prevents the color to rearrange the system spontaneously into a meson-meson state.  The   tunneling of heavy quarks is just  less probable.

Taking, as in Eq.~\eqref{delta} 
\begin{align*}
E=E_\alpha=E_\beta=m_1+m_2+\delta\,,
\end{align*}
with $\delta>0$ and $\delta\ll m_1,m_2$,
as suggested by experimental observation~\cite{Esposito:2016itg}, we get
\begin{align}
\Gamma\simeq \frac{2\pi^2}{R}\,\frac{(2\mu)^{3/2}}{\widetilde m(m_1+m_2)}\,\mathcal{T}\sqrt{\delta}\,,
\end{align}
with $R$ given in~\eqref{stima} and $\mu$ being the reduced mass of the meson-meson system.

For our purposes it is convenient to use the simple approximation  $m_1\simeq m_2=m$ so that, when the light quark tunnels through the barrier ($M=m_q$), we get~\footnote{The width of a particle of mass $E$ as in~\eqref{delta} decaying in two particles of mass $m$ would be $\Gamma\simeq m^{-3/2}\sqrt{\delta}$ using the same approximations. This  formula would give very different results for beauty and charm tetraquarks.}
\begin{align}
\Gamma\simeq \frac{\pi^2}{R}\frac{\sqrt{m}}{\widetilde m}\mathcal{T}\, \sqrt{\delta} \equiv A(m,\widetilde m,B,\ell)\sqrt{\delta}\label{width}\,,
\end{align}
where $A$ is the same parameter defined in~\cite{Esposito:2016itg}.

The masses in the prefactor $A(m,\widetilde m, B,\ell)$ change markedly  when passing from tetraquarks with beauty to charmed ones, retaining the same $m_q$.  We know however that data on $X$, $Z_c$ and $Z_b$ total widths  can be fitted very well  with the same  constant $A\simeq10$~MeV$^{1/2}$~\cite{Esposito:2016itg}.
 With a constituent light quark mass $m_q=308$~MeV and a  binding energy of $B_c=0.021$~GeV (see above), we obtain the fitted value for $A$ if the width of the barrier and, consequently, the total size of the state are
\begin{align}
\ell_c \simeq 1 \text{ fm}\,, \qquad R_c\simeq1.3\text{ fm}.
\end{align}
We also use here $\widetilde m_{c}(\equiv m_{[cq]})=1.98$~GeV~\cite{Chen:2015dig} and the average value $m_D=1.94$~GeV for $D$ mesons.
All these numerical values  are quite reasonable. The new thing is that $R_c\simeq 1.3$~fm suggests a separation in space of the two diquarks, making the state slightly larger than a standard hadron.

Is it reasonable to obtain the same value for $A$ in the beauty sector as well? Given the average $B$ meson mass $m_B=5.30$~GeV, the diquark mass $\widetilde m_b=5.32$~GeV~\cite{Ali:2011ug} and given $B_b\simeq B_c$ (see Eq.~\eqref{eq:binding}), one finds that this is possible provided that
\begin{align}
\ell_b\simeq0.8\,\ell_c\,,\qquad R_b\simeq1\text{ fm}\,.
\end{align}

The proportionality constant between $\Gamma$ and $\sqrt{\delta}$ can therefore be the same for charmed and beauty states, for the binding energies found above and if the beauty tetraquark is somewhat smaller than the charmed one, as it is  natural to expect.


\section{The $bb\bar b\bar b$ di-bottomonium}
The state searched by LHCb~\cite{Aaij:2018zrb} might present interesting consequences. If a $X,Z$-like state were to be discovered below the closest meson-meson threshold with given quantum numbers there would be two options: $i)$ a rather small negative binding energy with respect to threshold could lead to a hadron molecule interpretation, $ii)$ a negative binding energy of $10$~MeV or larger would speak in favor of a standard compact tetraquark, of the same kind as those above threshold discussed in the previous Sections. 

Consider the tetraquark $T=[\bar b\bar b]_{s=1}[bb]_{s=1}$,
where the diquark spins are fixed by symmetry.
There are no open beauty thresholds available for this state.  The short distance 
barrier picture described above would still suggest a rather narrow width for the decay $T\to \Upsilon(1S)\Upsilon(1S)^* \to 4\ell\,$,
which is the most accessible experimentally. Another possible decay is $T\to \eta_b(1S)\eta_b(1S)^*\to{\rm hadrons}$,
with in principle more phase space than the previous one, although way more challenging on the experimental side~\footnote{Diquarks are attractive in the antisymmetryc color antitriplet. Given the symmetry in flavor, $s=1$ is required. If $L=0$, overall $T$ could have total spin $J=2,1,0$. In the simplest case $J=0$ one can see that the probability to decay into $\eta_b\eta_b$ is $3/4$ and  $1/4$ to decay into $\Upsilon\Upsilon$. This would be a rather interesting prediction if only one could see $\eta_b$'s with the same eases with which $\Upsilon$'s are detected.  }. 

As explained in~\eqref{eq:mass}, the mass of this state will be
\begin{align}
m_T=2\widetilde m_{bb}-B-\kappa_{bb}\,.
\end{align}
Since the chromomagnetic couplings are inversely proportional to the quark masses~\cite{Georgi:1982jb}, $\kappa_{cq}\simeq67$~MeV and $\kappa_{bq}\simeq10$~MeV~\cite{Ali:2011ug}, we assume $\kappa_{bb}\simeq0$. We also notice that the mass of the charm diquark, $\widetilde m_c$, is some 50-100 MeV heavier than the same flavor meson, while $\widetilde m_b$ is only 30 MeV above it. We then make the further assumption $\widetilde m_{bb}\simeq m_\Upsilon$.

The typical energy scale of the problem is the mass of the beauty. Hence we compute $B$ determining the ground state in the same potential introduced in the previous section, but for $bb$  diquark masses. The binding energy is found to be $B=0.113$~GeV, thus predicting the mass of the $bb\bar b \bar b$ tetraquark around
\begin{align}
m_T\simeq 18.8 \text{ GeV}\,.
\end{align}
For this tetraquark we  allow for smaller sizes $\ell_b$ of the potential barrier.

Can this resonance be seen in the $\Upsilon\mu\mu$ final state given the current sensitivity~\cite{Aaij:2018zrb,slides}? 
The three-body decay rate is written as~\footnote{We checked that all our results are unchanged when the initial configuration is considered as a two particle state.}
\begin{align} \label{1to3}
\begin{split}
d\Gamma(T\to\Upsilon\mu\mu)=&\int_{s_\text{min}}^{s_\text{max}}ds\,BW(s)\sqrt{\frac{\bm{p}^2_{\Upsilon^*}+m_{\Upsilon}^2}{\bm{p}^2_{\Upsilon^*}+s}} \\
&\times\frac{d\Gamma(\Upsilon^*\to\mu\mu)}{\Gamma(\Upsilon^*\to\text{all})}\,d\Gamma(T\to\Upsilon\Upsilon^*)\,.
\end{split}
\end{align}
Here $s_\text{min}=4m_\mu^2$ and $s_\text{max}=(m_T-m_\Upsilon)^2$, and $BW(s)$ is the $\Upsilon$ Breit-Wigner distribution. The $T\to\Upsilon\Upsilon^*$ rate is computed as in Eq.~\eqref{1to2} using $\widetilde m=\widetilde m_{bb}$, $m_1=m_\Upsilon$ and $m_2=\sqrt{s}$. The decay is due to the tunneling of the beauty quark through the barrier.

The LHCb collaboration reported upper bounds on $S=\sigma(pp\to T)\mathcal{B}(T\to\Upsilon\mu\mu)\mathcal{B}(\Upsilon\to\mu\mu)$~\cite{Aaij:2018zrb}. Although the production cross section for the $bb\bar b\bar b$ is unknown, we can reasonably assume for it to be similar to the one for the production of a pair of $\Upsilon$, $\sigma(pp\to T)\simeq 69$~pb~\cite{Khachatryan:2016ydm}. One could also take the production cross section to be of the same order as that of the $X(3872)$, i.e. $\sigma(pp\to T)\simeq30$~nb~\cite{Esposito:2016noz,Bignamini:2009sk}. The $X$ is so far the only tetraquark observed in prompt production.

\begin{figure}
\centering
\includegraphics[width=0.43\textwidth]{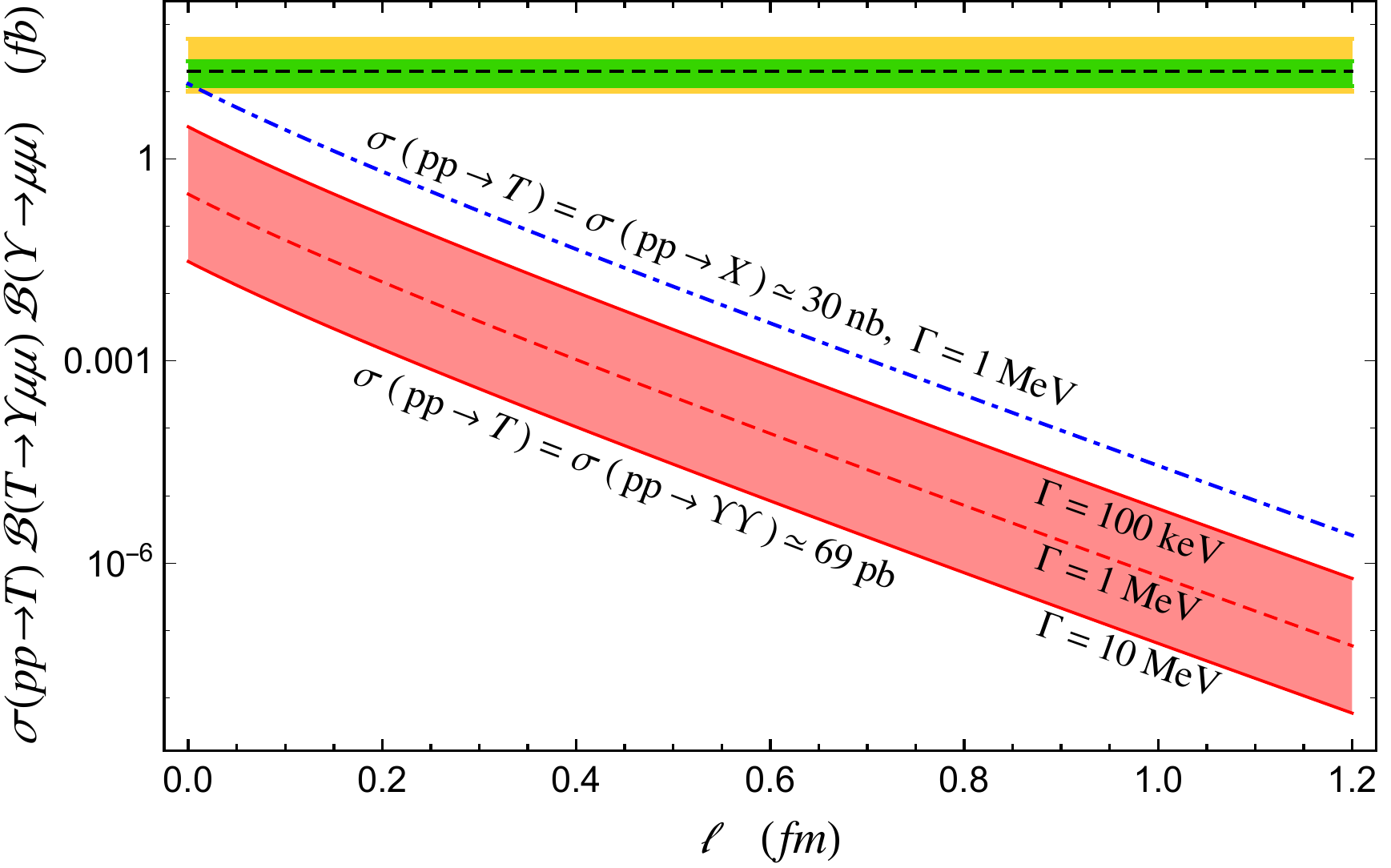}
\caption{Comparison between the experimental upper bounds presented in~\cite{Aaij:2018zrb} (black dashed line) and the predicted value of $S$ as a function of the size of the repulsive barrier. The green and yellow bands correspond respectively to the $1\sigma$ and $2\sigma$ C.L., while the red band spans different widths of the tetraquark. The red band corresponds to $\sigma(pp\to T)=69$~pb, while the blue dot-dashed line to $\sigma(pp\to T)=30$~nb. These results agree with~\cite{Karliner:2016zzc}. A state 500 MeV below threshold~\cite{slides} would have an even smaller cross section.} \label{fig:S}
\end{figure}

Integrating Eq.~\eqref{1to3} over the momenta of the final particles, we compute $S$ as a function of the size of the potential barrier and for a total width of the di-bottomonium between 1 keV and 10 MeV. In Fig.~\ref{fig:S} we compare it with the LHCb results. 
We find the partial width for the $T\to\Upsilon\mu\mu$ decay to be too small to be currently observed at the LHC.
If the $bb\bar b\bar b$ resonance were to be 500 MeV below threshold, as suggested by the preliminary CMS analysis~\cite{slides}, the decay rate would be further suppressed.

\section{Conclusions}
In this paper we propose a model to estimate masses and widths of tetraquarks based on  the conjecture of a short range diquark repulsion in a compact tetraquark. We assume that the two diquarks are separated by a potential barrier of size $\ell$, and that the decay occurs when one of the quarks tunnels through the barrier. The internal dynamics makes the tetraquarks slightly larger than the corresponding mesons.

This solves a number of open issues:
\begin{enumerate}
\item the spin-spin interactions are confined within the diquarks;
\item the tetraquark states decay more copiously into open flavor mesons rather than quarkonia;
\item the mass splitting between the first radial tetraquark excitation and its ground state is correctly reproduced assuming that the diquark potential has negligible Coulomb term~\cite{Chen:2015dig};
\item in~\cite{Esposito:2016itg} it was observed that tetraquarks follow the curve $\Gamma=A\sqrt{\delta}$ with the \emph{same} $A$ for both beauty and charm resonances.
\end{enumerate}

This picture allows to predict the mass of the di-bottomonium state $bb\bar b\bar b$, currently under experimental scrutiny~\cite{Aaij:2018zrb,slides}. 
Our model predicts a state approximately 100 MeV below threshold, but whose decay into $\Upsilon\mu\mu$ is unlikely to be observed at the LHC, as suggested by recent LHCb results~\cite{Aaij:2018zrb}. This dynamical picture also determines the effective coupling presented in~\cite{Eichten:2017ual,Eichten:2017ffp}.

\vspace{0.5em}

\begin{acknowledgments}

We are grateful to L.~Maiani, A.~Pilloni and S.~Rahatlou for valuable insights, and to R.~Barbieri, F.~Buccella and G.~Veneziano for short useful discussions. This research was supported in part by Perimeter Institute for Theoretical Physics. Research at Perimeter Institute is supported by the Government of Canada through Industry Canada and by the Province of Ontario through the Ministry of Economic Development \& Innovation. The work done by A.E. is partially supported by US Department of Energy grant de-sc0011941 and partially supported by the Swiss National Science Foundation under contract 200020-169696 and through the National Center of Competence in Research SwissMAP.

\end{acknowledgments}

\bibliographystyle{apsrev4-1}
\bibliography{biblio}

\end{document}